\documentclass[letterpaper, 10 pt, conference]{ieeeconf}  

\IEEEoverridecommandlockouts                              
\usepackage{amsmath} 
\usepackage{amssymb}  
\usepackage{xspace} 
\usepackage{algorithm,algorithmic}

\usepackage{cite}
\usepackage{xcolor}
\usepackage{graphicx}
\usepackage{subfigure}

\usepackage[hidelinks]{hyperref}
\DeclareFontFamily{U}{wncy}{}
\DeclareFontShape{U}{wncy}{m}{n}{<->wncyr10}{}
\DeclareSymbolFont{mcy}{U}{wncy}{m}{n}
\DeclareMathSymbol{\Sh}{\mathord}{mcy}{"58}

\newcommand{\ie}{i\/.\/e\/.,\/~}
\newcommand{\eg}{e\/.\/g\/.,\/~}
\newcommand{\cf}{cf\/.\/~}
\newcommand{\fig}{Fig\/.\/~}

\newcommand{\sect}{Sec\/.\/~}

\usepackage{fancyhdr}
\newcommand{\mytitle}{\textbf{Accepted final version.}
To appear in 57th IEEE Conference on Decision and Control

\copyright 2018 IEEE. Personal use of this material is permitted. Permission from IEEE must be obtained for all other uses, in any current or future media, including
reprinting/republishing this material for advertising or promotional purposes, creating new collective works, for resale or redistribution to servers or lists, or reuse of
any copyrighted component of this work in other works.}
\newtheorem{remark}{Remark}

\DeclareMathOperator*{\argmin}{arg\,min}

\newcommand{\mieds}{MIEDS\xspace}
\newcommand{\emieds}{sMIEDS\xspace}

\title{\LARGE \bf
Efficient Encoding of Dynamical Systems\\through Local Approximations
}

\author{ \parbox{3 in}{\centering Friedrich Solowjow*
        \thanks{*Equally contributing}\\
        Max Planck Society\\
        {\tt\small fsolowjow@tuebingen.mpg.de}}
        \hspace*{ 0.5 in}
        \parbox{3 in}{ \centering Arash Mehrjou*\\
       Max Planck Society \\
        {\tt\small amehrjou@tuebingen.mpg.de}}\\
        \parbox{3 in}{\centering Sebastian Trimpe\\
        Max Planck Society\\
        {\tt\small strimpe@tuebingen.mpg.de}}
}

\author{Friedrich Solowjow$^{1,*}$, Arash Mehrjou$^{2,*}$, Bernhard Sch{\"o}lkopf$^{2}$, Sebastian
Trimpe$^{1}$
\thanks{*Equally contributing.}
\thanks{$^{1}$Intelligent Control Systems Group, Max Planck Inst.\ for Intelligent Systems, 70569 Stuttgart, Germany. Email: \{fsolowjow, trimpe\}@is.mpg.de.}%
\thanks{$^{2}$Empirical Inference Department, Max Planck Inst.\ for Intelligent Systems, 72076 T\"{u}bingen, Germany. Email: \{amehrjou, bs\}@tue.mpg.de.}
\thanks{This work was supported in part by the Max Planck Society, the Cyber Valley Initiative, and the German Research Foundation (DFG) grant TR 1433/1-1. The authors thank the International Max Planck Research School for Intelligent Systems (IMPRS-IS)
for supporting Friedrich Solowjow.}}%

\begin{document}

\maketitle
\thispagestyle{fancy}
\pagestyle{empty}

\begin{abstract}

An efficient representation of observed data has many benefits in various domains of engineering and science. 
Representing static data sets, such as images, is a living branch in machine learning and eases downstream tasks, such as classification, regression, or decision making. However, the representation of dynamical systems has received less attention. 
In this work, we develop a method to represent a dynamical system efficiently as a combination of a state and a local model, which fulfills a criterion inspired by the minimum description length (MDL) principle. 
The MDL principle is used in machine learning and statistics to quantify the trade-off between the ability to explain seen data and the model complexity.
Networked control systems are a prominent example, where such a representation is beneficial. When many agents share a network, information exchange is costly and should thus happen only when necessary.
We empirically show the efficiency of the proposed encoding for several dynamical systems and demonstrate reduced communication for event-triggered state estimation problems.
\end{abstract}

\section{Introduction}
\label{sec:intro}
Nonlinear functions are essential for modeling complex physical systems. However, utilizing nonlinear models for practical applications is difficult and computationally expensive. Approximations are often required in order to obtain feasible algorithms.
In this paper, we propose a framework that adaptively chooses \emph{simple} local models, which are chosen as complicated as required to obtain a desired approximation accuracy. 
The intuition behind our approach goes back to Occam's razor principle, which states that among competing models with equal ability to explain seen data, the simplest model should be preferred. 
Based on Occam's razor principle, rigorous methods in machine learning have been established~\cite{barron1998minimum}.
One of them is the minimum description length (MDL) principle, which is frequently used to investigate the connection between compression and generalization, \cf \cite{vapnik2013nature}. 
The main idea of the MDL principle comes down to describing a given data set with the aid of a hypothesis or law. This way, the information content of the data set can be compressed. However, in order to achieve a benefit in terms of compression, the hypothesis also needs to be simple enough with respect to the data set. Therefore, the MDL-principle quantifies the trade-off between accuracy and complexity of a model and yields an efficient representation of data.
\begin{figure}
\includegraphics[width=0.48\textwidth]{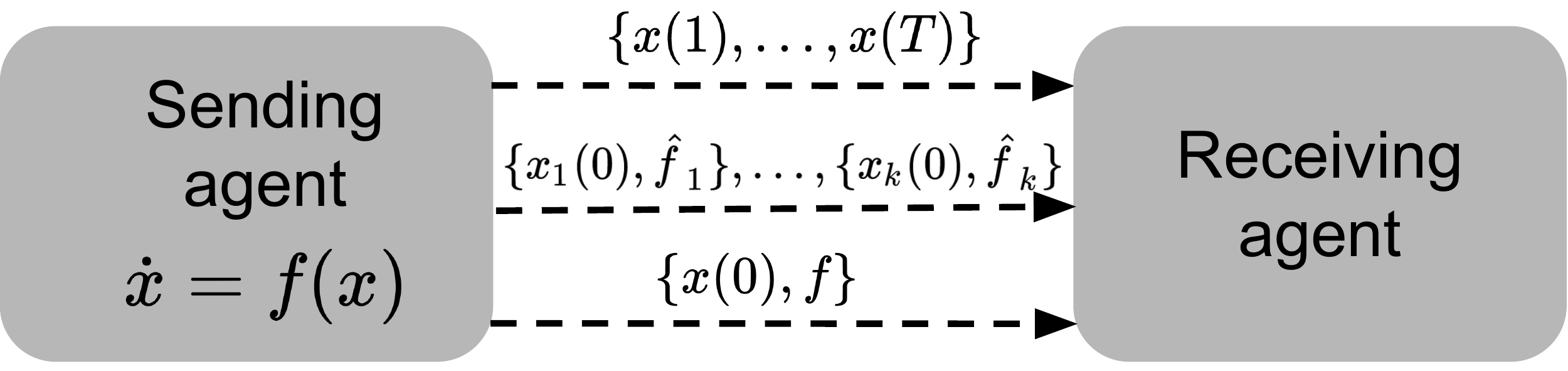}
\caption{Illustration of different encoding schemes of dynamical systems. Depicted are two agents, where one receiving agent requires the states of a sending agent.
There are many possibilities how to transmit information about the dynamical system. 
This can be state data itself $x(1),\ldots,x(T)$, or an initial point $x(0)$ and the dynamical function $f$, based on which it is possible to generate the whole trajectory of states. Since the dynamical function $f$ is an infinite dimensional object in general, it can be infeasible to transmit it.
Therefore, we develop a scheme that adaptively chooses local dynamics approximations $\hat{f}$, and propose this as an efficient encoding of dynamical system information.
\vspace{-0.2cm}}
\label{fig:block}
\end{figure}
Such an efficient encoding of dynamical systems has applications in various areas, such as statistical inference~\cite{ionides2006inference}, compression, and transmission~\cite{hamilton2013modelling}---basically, whenever dynamics or state information need to be transmitted or stored efficiently. 

\emph{Event-triggered state estimation} (ETSE) has been developed over the last decade \cite{YoTiSo02,TrDAn11,sijs2012event,TAC14b_web,wu2013event,Mi06}
for efficient transmission of data in networked control systems. 
In common ETSE approaches, dynamical system models are used to predict agents' behavior. This way, periodic transmission of state data can be avoided, and state updates are only transmitted whenever the model predictions are not sufficient. This stream of work has recently been extended in \cite{so18} to also consider communication of dynamical system models, in addition to states.
However, the framework in \cite{so18} has been developed for linear systems and the extension to nonlinear systems is nontrivial. This is due to the fact that the function $f$, describing dynamical system $\dot{x}(t) = f(x(t))$, is an infinite dimensional object and that it is not clear how to represent the information content of this system in general.
We assume that data from this system is available as a time-sampled signal, as is usually the case in practice. 
Therefore, there are two extremes in representing the history of the states $x(t)$ over a given time horizon $T$. 
One is storing or transmitting the states $x(t)$ for each time instance. The other one would be storing an initial point $x_{0}$ and the dynamical function $f$. 
For a deterministic system, the former can be produced by the latter, however the required storage space will be very different if the function $f$ can be represented efficiently. 
In this paper, we investigate an optimal middle point within this spectrum by finding a tractable representation of $f$. 
These different possibilities are depicted in \fig \ref{fig:block}, where two agents that exchange information are illustrated. Furthermore, the solution in between is also proposed, \ie local approximations $\hat{f}_1,\ldots,\hat{f}_k$.

For the scenario, where two agents transmit information as in \fig \ref{fig:block}, we seek to minimize overall communication between agents by transmitting local models in addition to state data. The MDL principle provides a structured approach to achieve this and yields
an unifying framework for discussing communication of state and model data in the context of ETSE. 
With the aid of the proposed algorithm, efficient and lighter models can be found that outperform common ETSE methods in terms of communication requirements. 
In particular, whenever the dynamics are obtained through learning methods, \eg Gaussian process regression~\cite{doerr2017optimizing}, or neural networks \cite{narendra1990identification}, these models can be huge, since they incorporate the whole training data set or rely on many parameters. Therefore, transmission of the learned function $f$ might be infeasible.
\paragraph*{Contributions}
In summary, this paper makes the following main contributions:
\begin{itemize}
\item Proposing a local encoding of dynamical systems, based on a MDL-type objective, which uses basis expansions to achieve adaptive model complexities (\eg the number of terms in Taylor expansion);
\item Based on this, the development of the \emph{Minimum Information Exchange in Distributed Systems} (\mieds) algorithm, combining two ideas (i) local approximations of the function $f$; and (ii) optimal choice for the accuracy of each local estimation;
\item Application to ETSE for nonlinear stochastic systems and thus extending the \mieds algorithm to stochastic problems;
\item Proposing a unifying framework for common ETSE methods.
\end{itemize}
We focus on the information exchange between one sending and one receiving agent. Extending our results to multiple agents is possible, but requires extra work and is beyond the scope of the current paper. 

\paragraph*{Related work}
Considering communication in networked control systems as a limited resource gave rise to various event-triggered communication architectures for estimation and control.
In particular, event-triggered state estimation and control are well studied for linear dynamical systems; for recent overviews see, \eg \cite{shi2015event,trimpeEBCCSP15,HeJoTa12,Mi15} and references therein. 

In this paper, we mainly focus on remote state estimation, as in \fig \ref{fig:block}, for nonlinear dynamical systems. There are some results in ETSE for nonlinear systems \cite{TrBu15,MaEsGaSa15}, however, they utilize a fixed global dynamical function $f$ in order to predict future behavior on a remote side. 
Furthermore, there is little work that considers the transmission of dynamical models in addition to states. For linear systems, this problem has been addressed in \cite{so18} together with quantifying expected communication rates.

The MDL principle is a practical realization of Occam's razor~\cite{barron1998minimum} and seeks for a model with a short description, which also yields compact and well behaved data.
This compression of data finds applications in various branches of statistics and machine learning such as statistical inference~\cite{balasubramanian1997statistical}, unsupervised learning~\cite{zemel1994minimum}, Bayesian model selection~\cite{mehrjou2016improved}, and causal inference~\cite{janzing2010causal}. The information content of dynamical systems is investigated in~\cite{balduzzi2008integrated, kleeman2011information}.

The benefits of approximating dynamics models through local representations have long been recognized \cite{atkeson1997locally, nelles1996basis}.  
While in these works, the order of the local models is usually fixed (e.g., linear), we propose herein to optimally choose the model complexity.  Thus, we typically obtain models of different order for different state space regions.
\section{Problem Formulation}
Consider a deterministic dynamical system
\begin{equation}
 \dot{x}(t) = f(x(t)),\quad x(t_0) = x_0,
\label{eq:sys}
\end{equation}
within a finite time horizon $t \in [t_0, t_0+T]$.
We assume $f$ is a known nonlinear, Lipschitz continuous function, and the states $x(t) \in \Omega \subset \mathbb{R}^{n}$ can be measured. 
The system \eqref{eq:sys} can also represent a controlled system by incorporating the effect of some state feedback $u(t)=g(x(t))$ within $f$. 

In this paper, we develop a tractable algorithm to encode the information of the dynamical system \eqref{eq:sys} efficiently. We assume that data from this system is available as a time-sampled signal $x(1), x(2), \ldots$. Furthermore, we want to point out that it is not possible to transmit the full stream of continuous data and mention the slight abuse of notation in the indexing here.

Therefore, one possibility to represent the considered system is by storing all states at every time instance, \ie $[x(1), x(2), \ldots, x(T)]$. Alternatively, it is also possible to store the function $f$ and an initial state $x(0)$. With the aid of the dynamical function, it is then possible to recover the trajectory of states and thus, data is compressed. 
However, it is in general not clear how to store a nonlinear function $f$ in an optimal way.
Therefore, we propose to approximate the function $f$ through $\hat{f}$ with the aid of a finite predefined basis $\phi_1,\ldots\phi_k$ (to be made precise later), which yields the system
\begin{equation}
\dot{\hat{x}}(t) = \hat{f}( \hat{x}(t) ) = \sum\limits_{i=0}^k \alpha_i \phi_i(\hat{x}),\quad \hat{x}(t_0) = x_0,
\label{eq:approx_sys}
\end{equation}
where the $\alpha_i$ are weights.
Since the dynamical function is used to recover a trajectory of states, approximating it by $\hat{f}$ introduces an error to the corresponding state sequence. Choosing an optimal complexity $k$ will be addressed in this paper. 
Furthermore, we propose to use local approximations $\hat{f}_1,\ldots,\hat{f}_m$, with varying complexities. The choice of the optimal number of local models $m$ will also be discussed later.

\section{The MDL Principle for Dynamical Systems}
\label{sec:mieds}
In this section, we introduce an MDL-type objective, which is the core of the proposed encoding scheme. In particular, it quantifies the trade-off between model complexity and accuracy. 
The resulting encoding scheme we denote as the \mieds algorithm.

{\it Minimum description length---} The minimum description length (MDL) principle is a formalization of the well investigated trade-off between model complexity and accuracy in explaining data. 
First, we will explain the MDL principle in general for a given data set $D= \{(z_1,g(z_1)),\ldots, (z_n,g(z_n)) \}$ and a hypothesis space of models or mechanisms $\mathcal{M}$. The goal is finding the best model $\hat{f} \in \mathcal{M}$, which explains the data as accurately as possible, while still being simple. 
We quantify this trade-off with the objective function
\begin{equation}
\label{eq:general_MDL}
\min\limits_{\hat{f} \in \mathcal{M}} \left( L_{\rm comp}(\hat{f}) + L_{\rm pred}(D|\hat{f})\right),
\end{equation}
where the first term $L_{\rm comp}$ is a measure of complexity of the function $\hat{f}$ and the second term $L_{\rm pred}$ is the prediction accuracy provided by this candidate function. 
As an example, assume a regression task, where we want to fit a polynomial function. In this case, $L_{\rm pred}$ measures the accuracy of the fit and $L_{\rm comp}$ is the order of the polynomial. 

For systems of the type \eqref{eq:sys}, the relevant data set is $D= \{(t_1,x_1),\ldots, (t_n,x_n) \}$.
The difference with the usual formalism of MDL is that instead of explicitly approximating the function of interest $x(t)$ by $\hat{x}(t)$, we approximate the right-hand side function $f$ of the dynamical system \eqref{eq:sys} by $\hat{f}$. This gives us an implicit approximation of $x(t)$, which is produced by simulating a dynamical system from $x(0)$ via $\hat{f}$ rather than $f$. This way, we efficiently compress information about the dynamical system \eqref{eq:sys}.

{\it The MDL-type objective functional---} In order to find the appropriate complexity of the local models, we use the MDL principle and propose to optimize the following functional
\begin{equation}
\min\limits_{k \in \{1,\dots,k_{\rm max}\}} \lambda k + \int_0^T \| x(t) - \hat{x}(t) \|_2 \,{\mathrm{d}}t,
\label{eq:mdlmodelselection}
\end{equation}
which is a concrete realization of \eqref{eq:general_MDL}.
Here, $\hat{x}(t)$ is defined as in \eqref{eq:approx_sys}, and $k$ is the number of terms we use for the approximation, which directly influences the accuracy of $\hat{x}(t)$. We only need to store or communicate the weights $\alpha$ in front of the basis functions $\phi$, since these are assumed fixed a priori. The parameter $\lambda > 0$ can influence the relative weight of the model complexity. Together with the partitioning of the state space, this affects what kind of functions are preferred by the algorithm.

{\it Intuition behind local models---} What we eventually want to store, is the history of states of a dynamical system. For many practical systems, the states remain in a small subset of the entire state space, which makes the local approach to encoding powerful. Therefore, we care about a local representation of a function based on a finite set of basis functions
where $\hat{f}$ is the local approximation to $f$ around a working point $x^*$. The set $\{\phi_i\}$ is chosen with cardinality $k_{\rm max}$ to contain basis functions that are able to approximate $f$ arbitrarily well as $k_{\rm max}\to \infty$. There are many well known types of basis functions and expansions that could be used such as Taylor series, Fourier series, or Legendre polynomials \cite{alt1992linear}. In this paper, we use Taylor expansions to showcase our points and thus assume $f$ is smooth enough.
\begin{figure}
\includegraphics[width=0.48\textwidth]{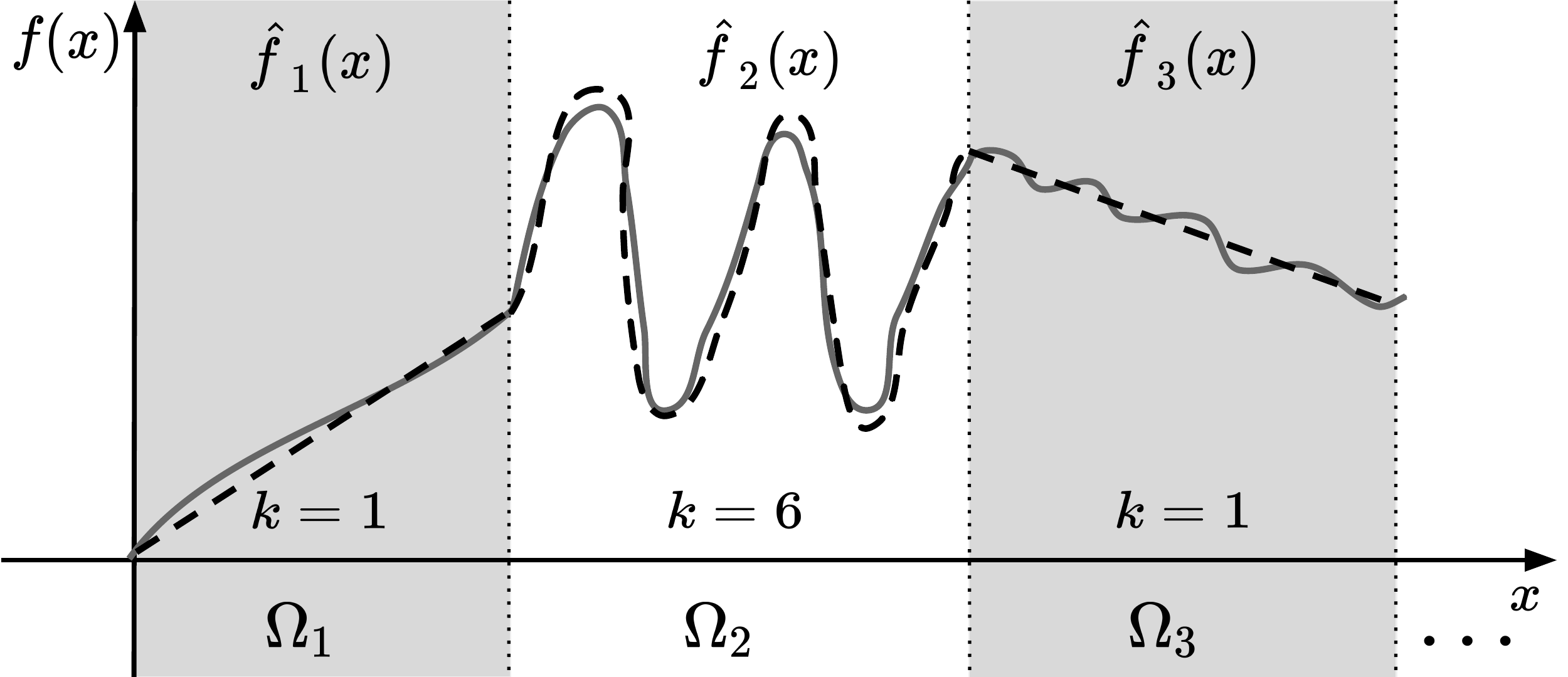}
\caption{Illustration of the proposed encoding for dynamical functions. 
The real dynamics (solid line) and local approximations (dashed line) are depicted.
The state space is partitioned into disjoint sets $\Omega_i$, in which the corresponding local approximation $\hat{f}_i$ is valid. 
Based on a minimum description length type objective the local complexity order $k$ of a Taylor expansion is determined.
While the graphs are sketched to highlight key aspects of the approach, the same properties are observed in the numerical examples herein.
\vspace{-0.2cm}}
\label{fig:schematics}
\end{figure}

The information content of the full function $f$ can be distributed to many local approximations.
One advantage of local models is the ability to sample them fast. When facing demanding and computationally expensive control or filtering problems, this might be a key aspect to achieve online applicability.
Furthermore, due to physical limitations or restrictions, the states $x(t)$ may stay within some bounded domain. Hence, the potential global input domain $\Omega$ reduces to the \emph{effective} input space, which we eventually care about. 

Figure \ref{fig:schematics} sketches the main idea behind the local approximations.
In the $\Omega_1$ part of the state space, the function is almost linear and hence, a linear approximation is used ($k=1$). In $\Omega_2$, however, there are significant nonlinearties and therefore, a polynomial with degree $k=6$ is chosen. A different effect can be seen in $\Omega_3$, where a linear model is preferred, because here, a marginal improvement in the states does not justify the cost caused by a more complex model. This trade-off, which is quantified in \eqref{eq:mdlmodelselection}, is essential in order to determine automatically the required model complexity.

{\it Partitioning and local models---} We use an equidistant partitioning of the state space, and the number of necessary partitions is determined through the proposed algorithm. Furthermore, we require that the state space is bounded. However, both assumptions could be relaxed by checking the distance between $x(t)$ and $\hat{x}(t)$ and refining the partitions on necessity. Applying other more sophisticated methods is also possible here.

Here, we define the maximal number of partitions $m_\mathrm{max}$ as a hyperparameter, which should be chosen large enough. We propose to partition the time domain and thus $T_{\rm local}=\frac{T}{m}$. Clearly, the local time horizon gets shorter as the number of partitions grows. We iterate over all possible choices of $m$ by locally solving for the optimal value of the cost functional \eqref{eq:mdlmodelselection} and summing over all local costs, which yields
\begin{equation}
	\argmin\limits_{m \in \{1,\ldots,m_{\rm max}\}} L_{\rm total}(m)=\sum_{i=1}^m L_i(k_i^*).
\end{equation}
The local optimal complexity $k_i^*$ is obtained by optimizing each individual local cost over the time horizon $T_{\rm local}=t^{\rm stop}_i-t^{\rm start}_i$. Therefore, we obtain
\begin{equation}
L_i(k_i) = \lambda k_i + \int_{t^{\rm start}_i}^{t^{\rm stop}_i} \| x(t) - \hat{x}(t)\|_2 \,{\mathrm{d}}t. 
\label{eq:mdlmodelselection_local}
\end{equation}

\begin{remark}
\label{rem:partitioning}
The local times also induce a partitioning in space, which we denote with $\Omega_i$. We obtain the $\Omega_i$ by evaluating the state trajectory at the desired switching point in time. It is also possible to define the partitioning in the state space and afterwards compute the corresponding times.
\end{remark}
To summarize, the complete encoding scheme consists of fixing a partitioning of the state space and finding optimal local approximations with respect to the introduced MDL-type objective functional \eqref{eq:mdlmodelselection_local}.
Afterwards, we change the partitioning, \eg increase $m$, and optimize again the local cost. Finally, we obtain the optimal $m^*$ and the corresponding models $\hat{f}_1,\ldots,\hat{f}_m$ with varying complexities $k_1^*,\ldots,k_m^*$. We call this the Minimum Information Exchange in Distributed Systems (MIEDS) algorithms due to its motivation from information exchange between distributed agents (\fig \ref{fig:block}).
In the next section, we demonstrate nontrivial behavior of $m$ and $k$ in numerical examples.
 \section{Experiments: Encoding}
We provide numerical experiments to demonstrate different aspects of the \mieds algorithm in encoding a deterministic dynamical system.

\begin{figure}[t!]
	\centering
\subfigure[$m=1$]{
\includegraphics[width=0.45\linewidth]{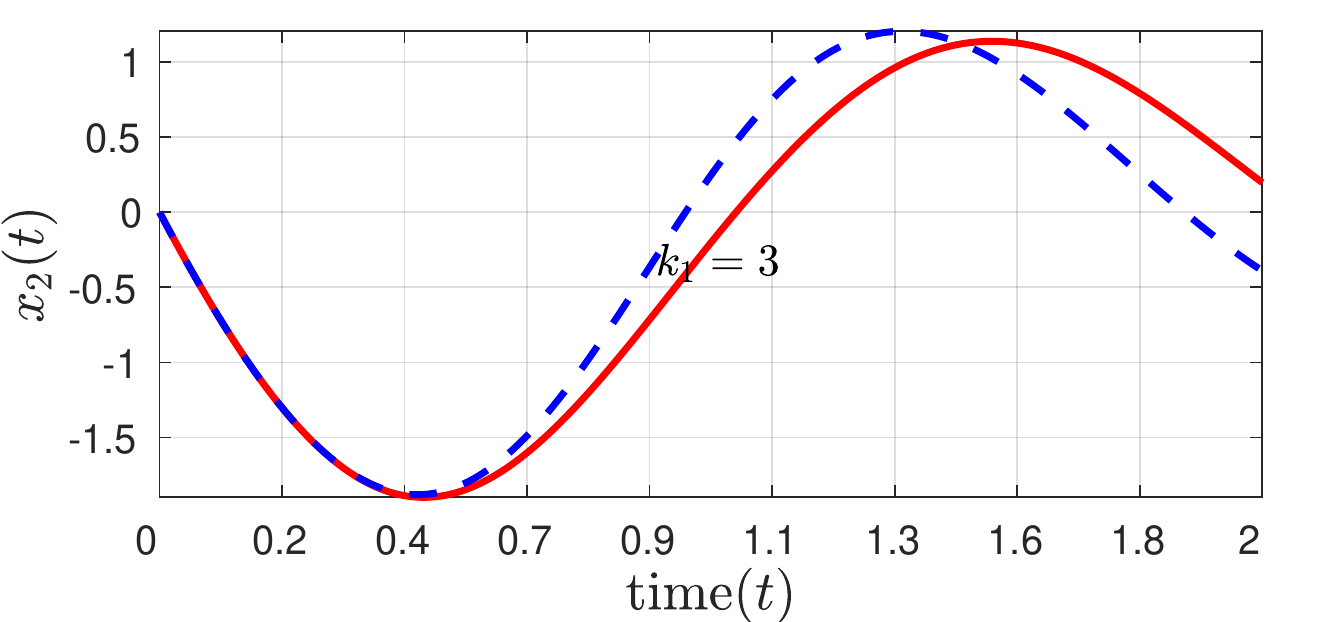}}
\subfigure[$m=2$]{
\includegraphics[width=0.45\linewidth]{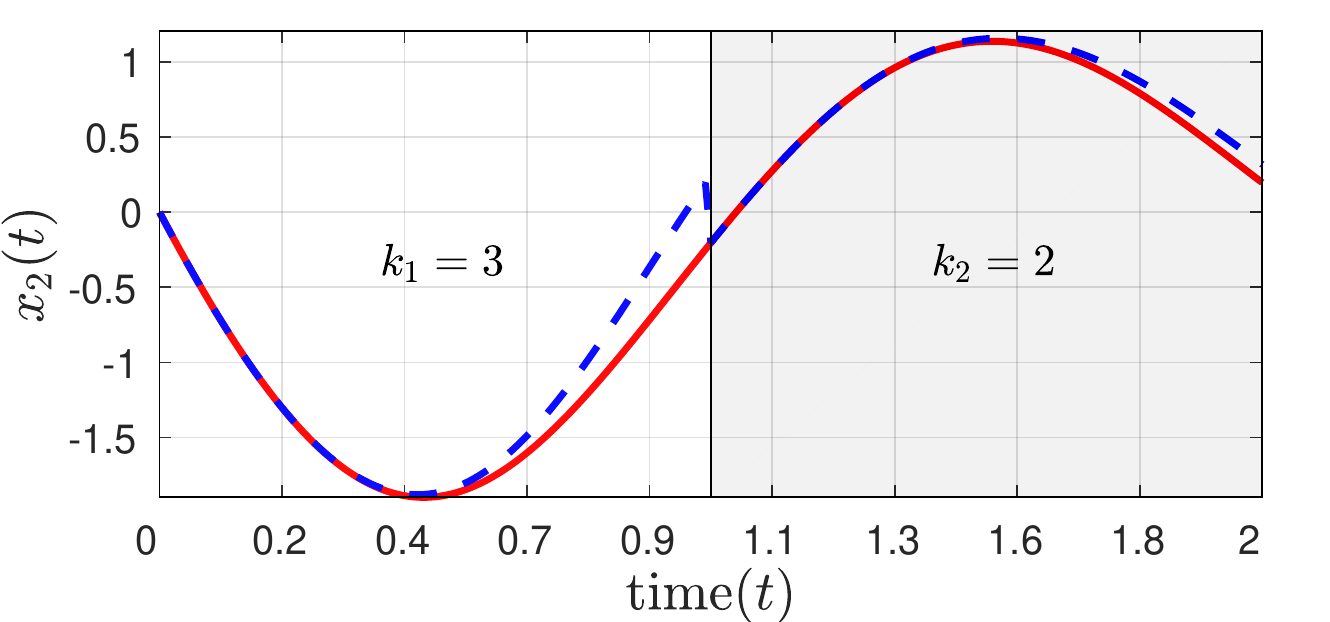}}
\subfigure[$m=3$]{
\includegraphics[width=0.45\linewidth]{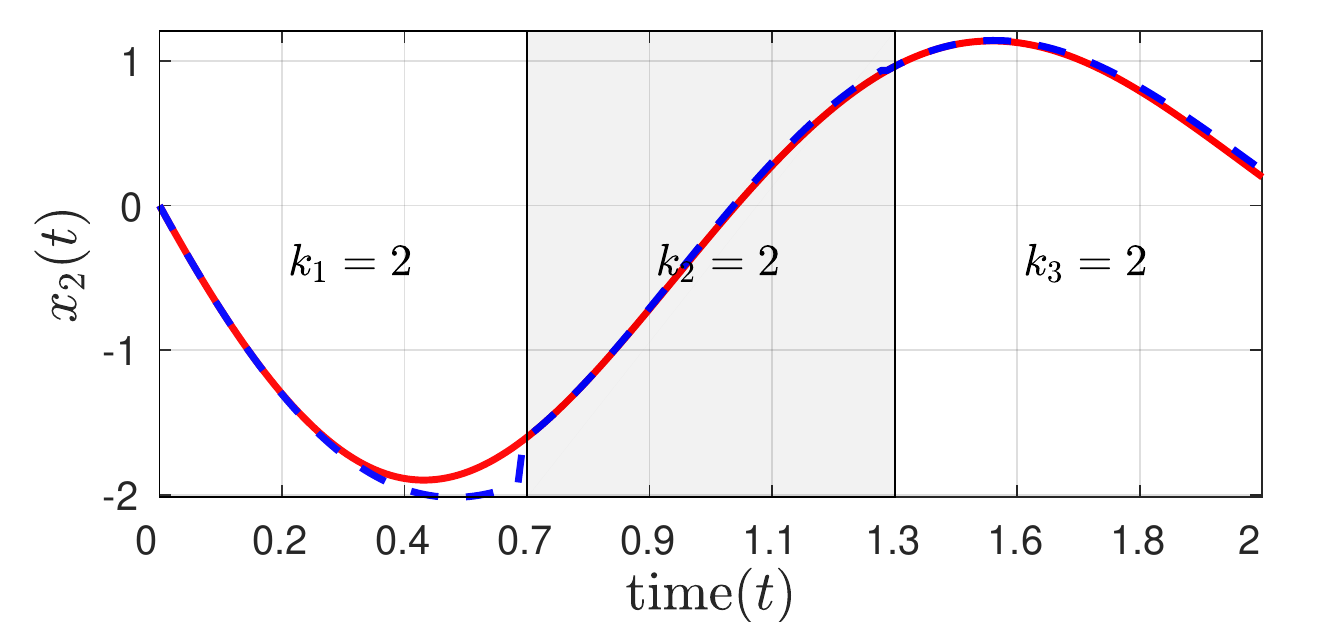}}
\subfigure[$m=4$]{
\includegraphics[width=0.45\linewidth]{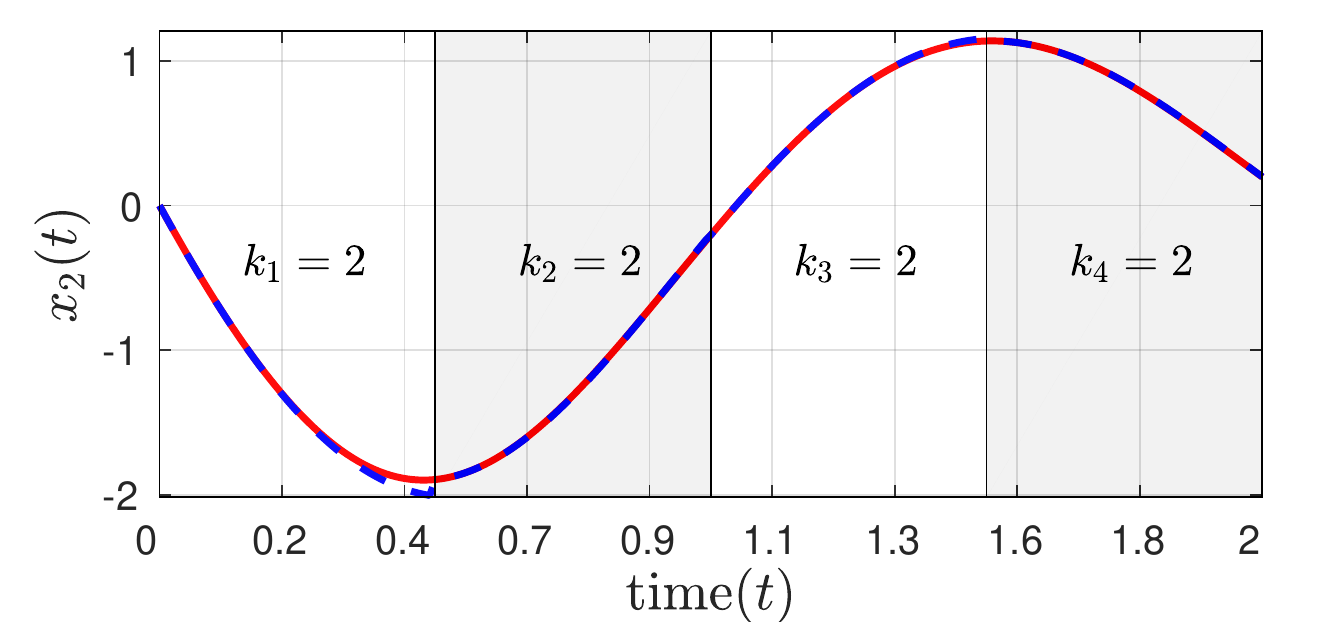}}\\
\subfigure[Cost function]{
\includegraphics[width=0.9\linewidth]{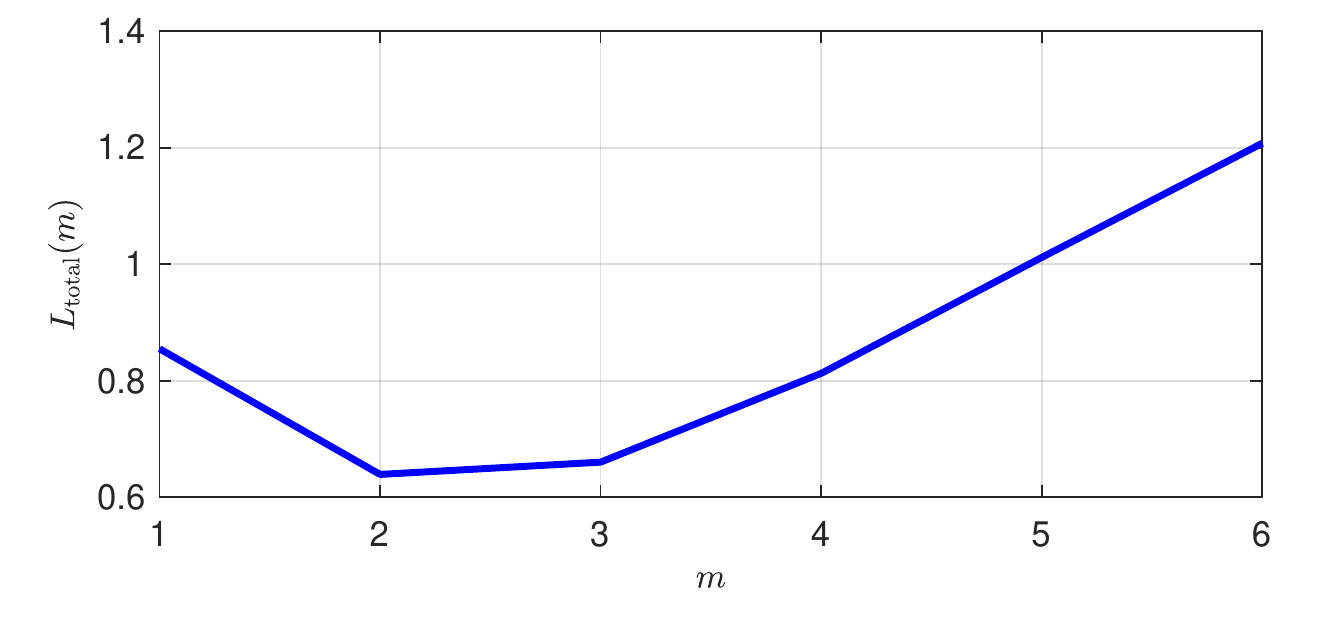}}
\caption{Pendulum dynamics. The top four figures show the system trajectory of the state $x_2(t)$ in red and different approximations resulting from our method in blue. Depending on the used partitions $m$ and Taylor terms $k$ we can observe different behavior.
The depicted cost function in Panel (e) shows a nontrivial optimal value $m^*=2$.\vspace{-0.2cm}}
\label{fig:pendulum}
\end{figure}

{\it Pendulum dynamics---} To start with, we consider the dynamics of a pendulum with two state equations $\dot{x}_1 = x_2$ and $\dot{x}_2 = -x_2 - 9.81{\rm sin}(x_1)$, where $x_1$ is the angular position and $x_2$ the angular velocity. The system starts from the initial state $[x_1(0),x_2(0)]=[\frac{\pi}{4}, 0]$. 
Assume polynomials with degrees up to $3$ are allowed as the set of basis functions to approximate each local region. Given this set, the approximation is done by Taylor expansions. The optimal degree $k_i$ of the Taylor expansion corresponding to each local region can be seen in \fig \ref{fig:pendulum}. We choose the parameters $\lambda=2$, $T=2$, and $k_{\rm max}=3$. The sampling time step $\Delta T$ is set to $0.01$.

{\it Quadrotor dynamics---} One typical example where dynamical system information is transmitted between agents, are Unmanned Aerial Vehicles (UAVs), whose states are constantly measured on-board, but only occasionally transmitted to the ground base. Continuous transmission of data is expensive in terms of battery power and bandwidth. 
To show the performance of \mieds on more complex dynamics, we present a dynamical system of a quad-rotor UAV 
\cite{sabatino2015quadrotor}.

In the following state space equation, $\phi\in[-\pi, \pi]$ is the roll (rotation around x axis) and $\theta\in(-\frac{\pi}{2},\frac{\pi}{2})$ is the pitch (rotation around y axis). The value of angles can violate the bounds but its interpretation is circular. The vector $[u, v, w, p, q, r]$ contains the linear and angular velocities in the body frame. The vector $[f_{wx}, f_{wy}, f_{wz}, f_t]=[1, 1, 1, 0]$ contains the wind forces and time-varying disturbance, $[\tau_{wx}, \tau_{wy}, \tau_{wz}]=[1, 1, 1]$ wind torques, and $[\tau_x, \tau_y, \tau_z]=[1, 1, 1]$ the control torques generated by the differences in the rotor speeds. We assume a mass $m=1$ and set the gravity constant to $g = 9.81$. This yields the system
\begin{equation}
\left\{
	\begin{array}{ll}
		\dot{\phi}=p+r(\cos(\phi)\tan(\theta))+q(\sin(\phi)\tan(\theta)) \\
        \dot{\theta}=q(\cos(\theta))-r(\sin(\phi))\\
        \dot{p}=\frac{I_y-I_z}{I_x}rq+\frac{\tau_x+\tau_{wx}}
        {I_x}\\
        \dot{q}=\frac{I_z-I_x}{I_y}pr+\frac{\tau_y+\tau_{wy}}
        {I_y}\\
        \dot{r}=\frac{I_x-I_y}{I_z}pq+\frac{\tau_z+\tau_{wz}}{I_z}\\
        \dot{u}=rv-qw-g\sin(\theta)+\frac{f_{wx}}{m}\\
        \dot{v}=pw-ru+g\sin(\phi)\cos(\theta)+\frac{f_{wy}}{m}\\
        \dot{w}=qu-pv+g\cos(\theta)\cos(\phi)+\frac{f_{wz}-f_t}{m}\\   
	\end{array}
\right.
\end{equation}
with an initial state $[{-2},{-3},1,3,1,4,2,1]$.

\begin{figure}[t!]
	\centering
	\hspace{-0.3cm}
\subfigure[roll $\phi$]{
\includegraphics[width=0.45\linewidth]{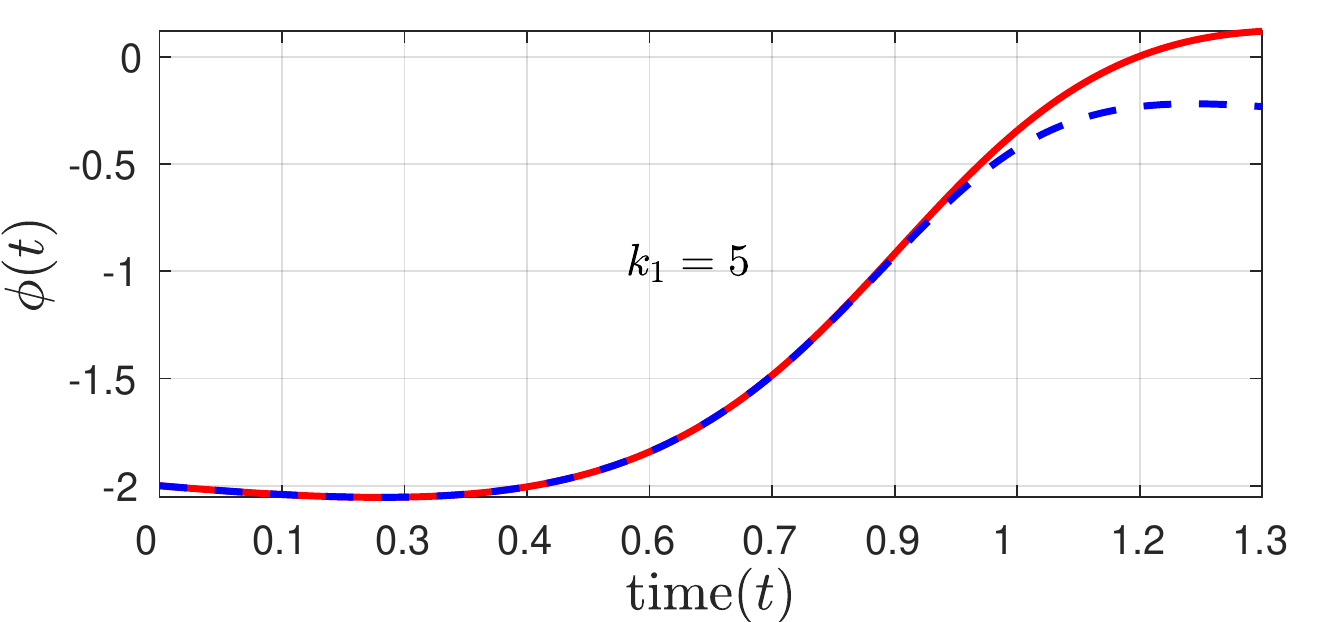}}
\subfigure[roll $\phi$]{
\includegraphics[width=0.45\linewidth]{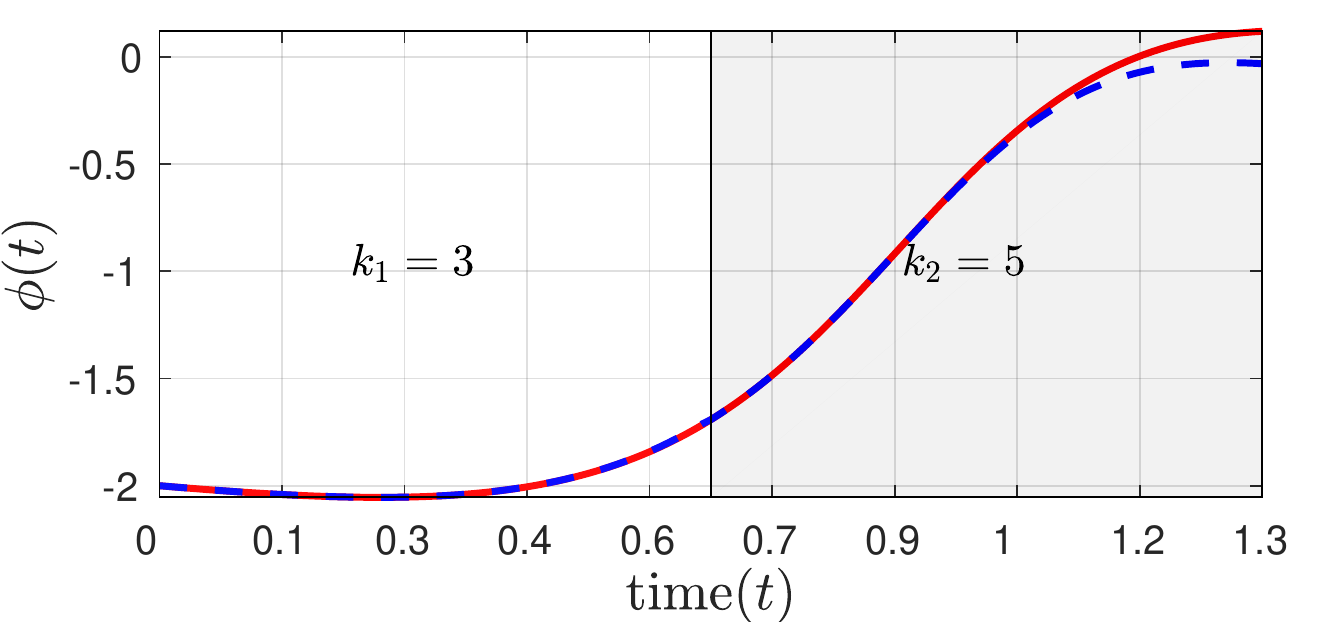}}\\
\subfigure[velocity $p$]{
\includegraphics[width=0.45\linewidth]{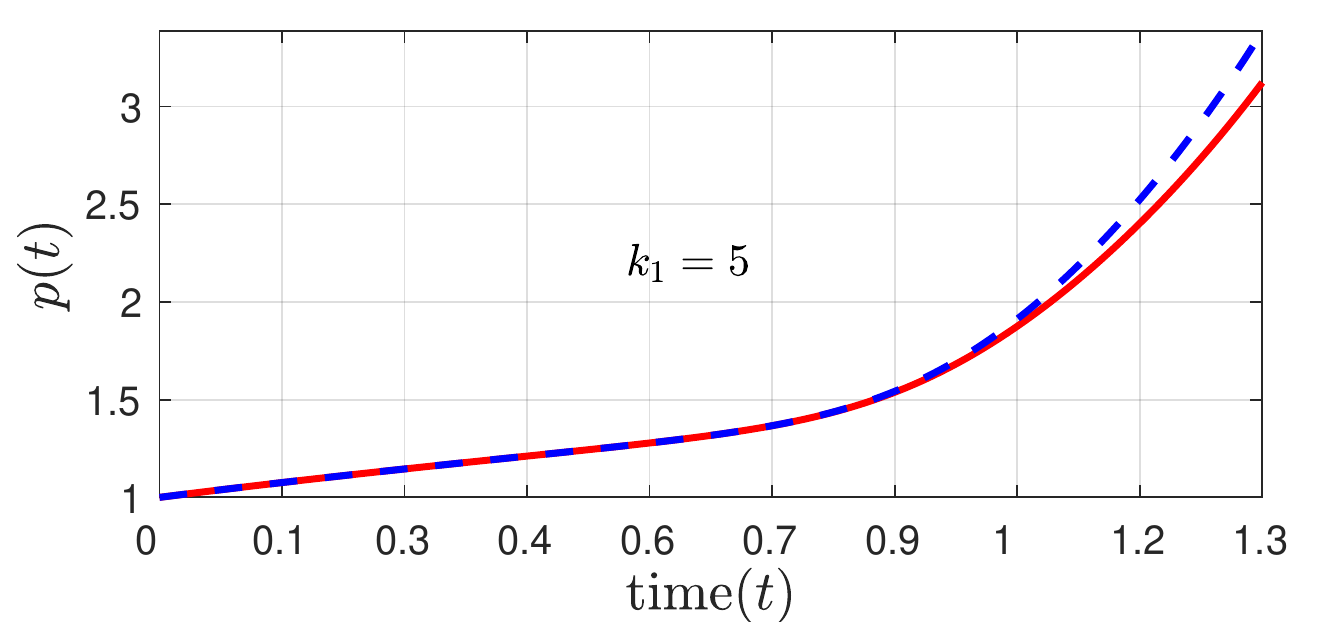}}
\subfigure[velocity $p$]{
\includegraphics[width=0.45\linewidth]{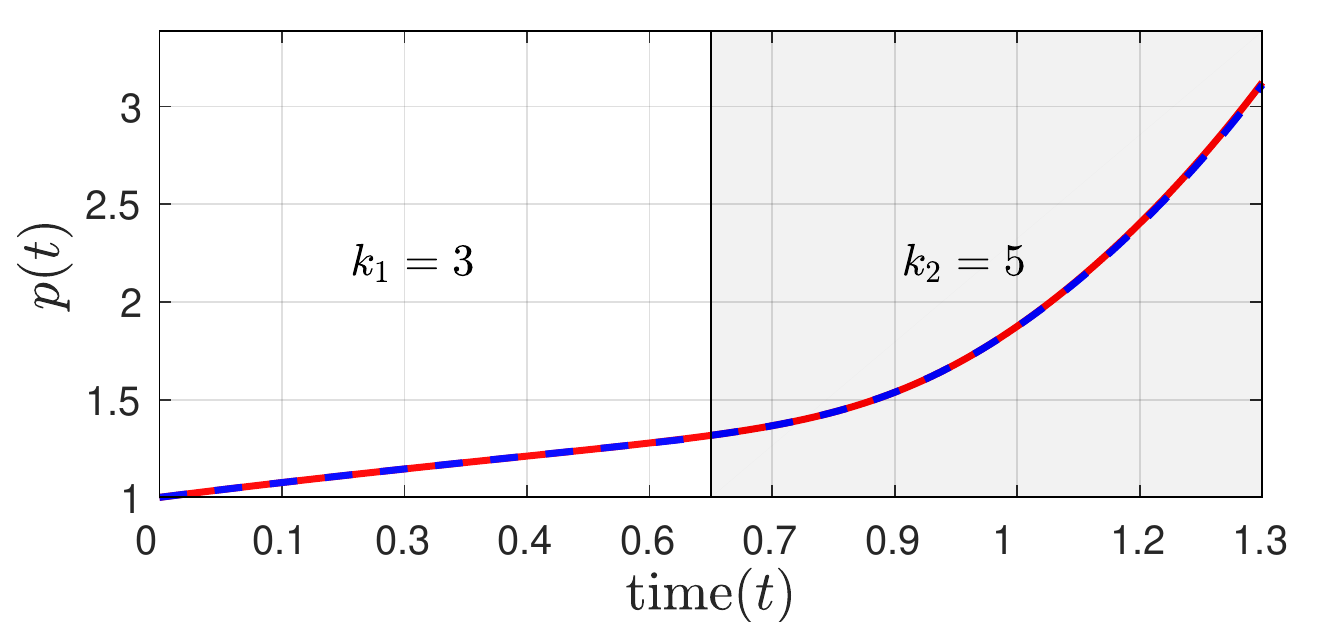}}
\subfigure[Cost function]{
\includegraphics[width=0.9\linewidth]{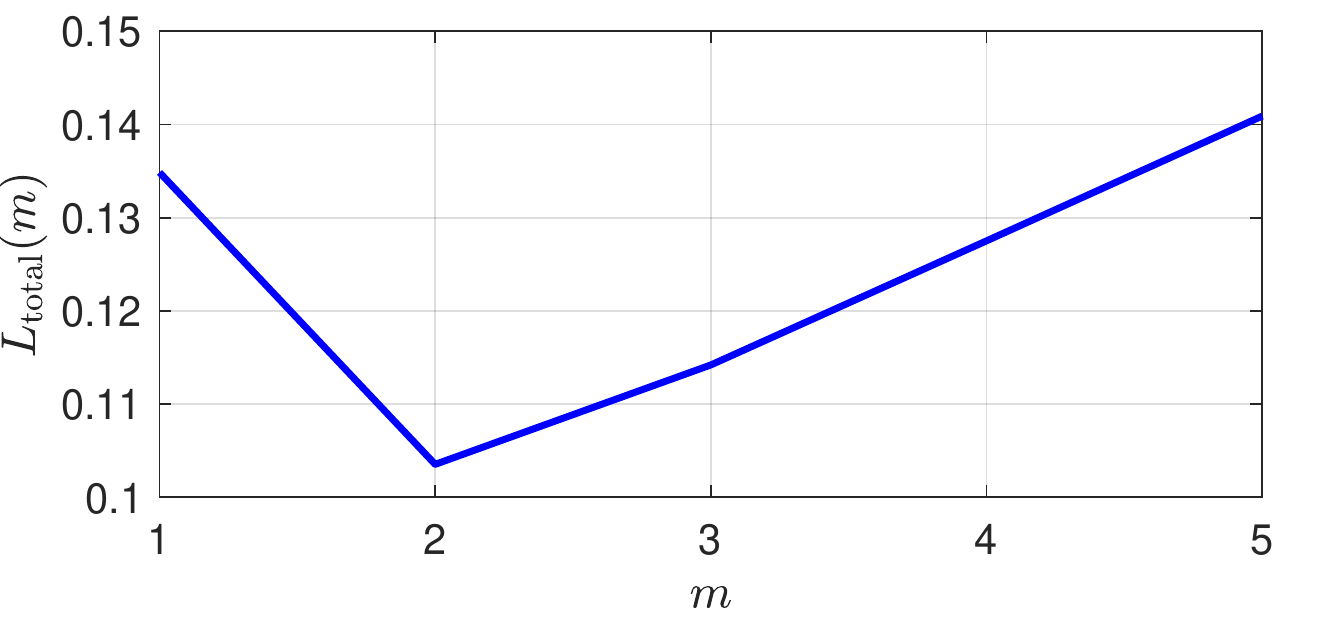}}
	\caption{Quadrotor dynamics. Top four figures show the effect of suboptimal ($m=1$, left) and optimal number of partitions ($m=2$, right) on two different states of the system. Having a single partition leads to complex models. Partitioning the state space into two local regions gives a fair trade-off between model complexity and data fit. The diagram (e) shows again a nontrivial optimum.\vspace{-0.2cm}}
	\vspace{-0.2cm}
\label{fig:quadrotor}
\end{figure}

In \fig \ref{fig:quadrotor}, we can see the effects of different model complexities on the state prediction accuracy. We take polynomials up to degree $5$ as allowed basis functions and the maximum number of partitions is upper bounded by $5$. Hence $k_{\rm max}=5$ and $m_{\rm max}=5$. The parameters $\lambda,T,$ and $\Delta T$ are chosen as before.
 
\section{Application to ETSE}
\label{sec:ETSE}
In this section, we demonstrate the effectiveness of our encoding scheme in ETSE scenarios, as motivated in the introduction. 
Because ETSE approaches typically consider stochastic systems, we first generalize \mieds to the stochastic setting (called stochastic \mieds (\emieds)).
\subsection{Extension to Stochastic Systems}
We consider stochastic dynamical systems of the type
\begin{equation}
 \dot{x}(t) = f(x(t)) + \epsilon(t),\quad x(t_0) = x_0,
 \label{eq:predictions}
\end{equation}
where $\epsilon$ is white noise and $f$ is a nonlinear function, which satisfies the conditions for the existence and uniqueness of solutions to stochastic differential equations (see \eg \cite{schuss2009theory}).

We adjust Equation \ref{eq:mdlmodelselection_local} by introducing a new variable $\bar{x}(t)$, which solves the deterministic problem without noise, which yields the local cost
\begin{equation}
      L_i(k_i) = \lambda k_i + \int_{t^{\rm start}_i}^{t^{\rm stop}_i} \| \bar{x}(t) - \hat{x}(t)\|_2 \,{\mathrm{d}}t.
\end{equation}
The partitioning should depend on the deterministic effects that drive the system and not on noise effects. Furthermore, the stochastic realizations vary between roll outs and we care more about the average behaviour of the system. 
\subsection{Event-triggered state estimation}
We introduce a typical ETSE framework, in order to apply our encoding scheme to it afterwards.
Consider the problem in \fig \ref{fig:block}, where we have two agents, one sending $v_{\rm send}$ and one receiving $v_{\rm receive}$.
We assume the behavior of agent $v_{\rm send}$ is described by \eqref{eq:predictions}. At the same time, the remote agent $v_{\rm receive}$ needs information about those states for tasks such as monitoring or control. We assume it is possible, but costly to communicate states through a network. Instead of transmitting data periodically, the receiving agent $v_{\rm receive}$ utilizes a model to obtain approximations $\hat{x}(t)$ of the true states $x(t)$. 
In order to ensure sufficient accuracy, we deploy the trigger $\gamma_{\rm noise} \in \{0,1\}$,
\begin{equation}
\gamma_{\rm noise}=1 \quad \Leftrightarrow \quad \| \hat{x}(t^*) - x(t^*) \| \geq \delta_{\rm noise}.
\label{eq:noisetrigger}
\end{equation}
If the trigger fires ($\gamma_{\rm noise} = 1$), $v_{\rm send}$ transmits the current state, and $v_{\rm receive}$ resets the prediction, $\hat{x}(t^*) \leftarrow x(t^*)$ and $t^*$ denotes the corresponding point in time.
The sending agent $v_{\rm send}$ implements \eqref{eq:noisetrigger} and thus also runs the same computations as $v_{\rm receive}$ to obtain $\hat{x}(t)$.
Due to the stochastic nature of the system, predictions can go arbitrarily bad, but \eqref{eq:noisetrigger} enforces a bound on the error. This is a common scheme in ETSE (cf.\ \emph{Related work} in Sec.\ \ref{sec:intro}).

Beginning from the most recently seen state $x(t^*)$, the receiving agent $v_{\rm receive}$ computes
\begin{equation}
\hat{x}(t) = \mathbb{E}\left[x(t) | x(t^*) \right], \quad t > t^*,
\label{eq:nlfiltering}
\end{equation}
where $\hat{x}(t^*) = x(t^*)$. 
Most ETSE schemes utilize the dynamics function $f$ in order to obtain $\hat{x}(t)$. For linear systems with Gaussian noise, the predictions \eqref{eq:nlfiltering} can be obtained analytically. However, for nonlinear systems, the computation is usually intractable and approximations are required.  
Here, we implement the approximation scheme
\begin{equation}
\dot{\hat{x}}(t) = \hat{f}(\hat{x}(t)), \quad \hat{x}(t_0) = x(t^*),
\label{eq:hatx}
\end{equation} 
which is not optimal, but straightforward and directly illustrates the importance of good models $\hat{f}$.
As an alternative to \eqref{eq:hatx}, particle filter type methods can be used, to obtain an unbiased estimate. However, these methods suffer immensely from inapt model representations, since they require to sample from the model repeatedly. Efficient local models help here a lot, in particular when we consider time sensitive online applications.

The \mieds algorithm yields a natural way to consider local models as transmittable data as well and yields, due to the MDL approach, a principled way of choosing optimal model complexities. We propose to transmit new models in an event-triggered fashion and thus reduce overall communication.

\subsection{Design of dual triggering scheme}
Next, we describe how to decouple stochastic effects from deterministic model errors, due to local approximations.
To integrate the \mieds algorithm into ETSE, we
propose to consider two triggers -- one to capture local random fluctuations $\gamma_{\rm noise}$ \eqref{eq:noisetrigger}, and one to cope with the change in the local dynamics $\gamma_{\rm dynamics}$. Whenever the trigger $\gamma_{\rm noise}$ fires, we communicate solely the current state. However, the trigger $\gamma_{\rm dynamics}$ is responsible to detect model inaccuracies, because of entering a new region $\Omega_i$ and communicate new local models when required. 

The trigger $\gamma_{\rm dynamics}$ is derived from applying the \mieds algorithm to the deterministic system $\dot{\bar{x}}(t) = f(\bar{x}(t)).$
This way, we obtain local models $\hat{f}^i$, each with a local domain $\Omega_i$. 
Assume $\bar{t}$ corresponds to $T_\mathrm{local}$ and is the point in time when switching should happen.
By evaluating $x(\bar{t})$, we obtain a switching point and illustrate this in \sect \ref{sec:experiments} for a one dimensional example.

Instead of switching the models after a predefined time, we emphasize the event triggered nature of our approach. Due to the noise, the system might stay arbitrary long within a local domain and the state dependent model switching is more robust to noise.
The trigger $\gamma_{\rm dynamics} \in \{0, 1\}$ fires whenever
\begin{equation}
x(t) \notin \Omega_i \quad \Leftrightarrow \quad \gamma_{\rm dynamics}=1.
\end{equation}
Hence, the local models are actually switched when necessary, based on the current state and not the current time. Depending on the application, it may be necessary to trigger on $\hat{x}(t)$ as well, or shrink the domain $\Omega_i$, since we know that $x(t)$ and $\hat{x}(t)$ are at most $\delta_{\rm noise}$ apart from each other.
By running the two trigger $\gamma_{\rm noise}$ and $\gamma_{\rm dynamics}$ in parallel, we obtain the stochastic \mieds (\emieds) algorithm, which utilizes the advantages of local models in a stochastic setting.

\subsection{Interpretation of existing ETSE schemes}
Standard event-triggered schemes can be understood in the sense of the MDL principle as well. 
Most protocols trigger communication whenever
$ \| x(t) - \hat{x}(t) \| \geq \delta$,
however, the computation of $\hat{x}(t)$ may differ for different strategies.
The send-on-delta (SoD) concept uses the latest communicated state, which means $\hat{x}(t^* ) = x(t^*)$. This coincides with predictions with the identity and hence, corresponds to the hypothesis space $\mathcal{M} = \{ {\rm Id} \}$.
For linear dynamics $f(x) = Ax$, we obtain $\mathcal{M} = \{ A : A \in \mathbb{R}^{n\times n} \}$. Usually, a fixed model is utilized to run the predictions, however, there are also approaches, where the dynamical model is actually chosen from that set through linear regression, as in \cite{so18}. 

With the above interpretation, the MIEDS formalism allows us to extend ETSE to the nonlinear setting.
For nonlinear dynamical systems, we assume the hypothesis space is given by a basis set of a suitable function space, \eg a power series or Fourier series. 
Hence, we obtain $\mathcal{M} = \{ \hat{f} : \hat{f}(x) = \sum_{i=0}^k \alpha_i \phi(x) \}$, where $\alpha_i$ are coefficients and $\phi_i$ basis functions. 
By communicating models in addition to states, we impose additional costs
in the ETSE setting. 

\section{Experiments: Reduced Communication}
\label{sec:experiments}
In this section, we present numerical experiments to demonstrate reduced communication behavior in an ETSE setting. 
\begin{figure}[tb]
\centering
\includegraphics[width=0.4\textwidth]{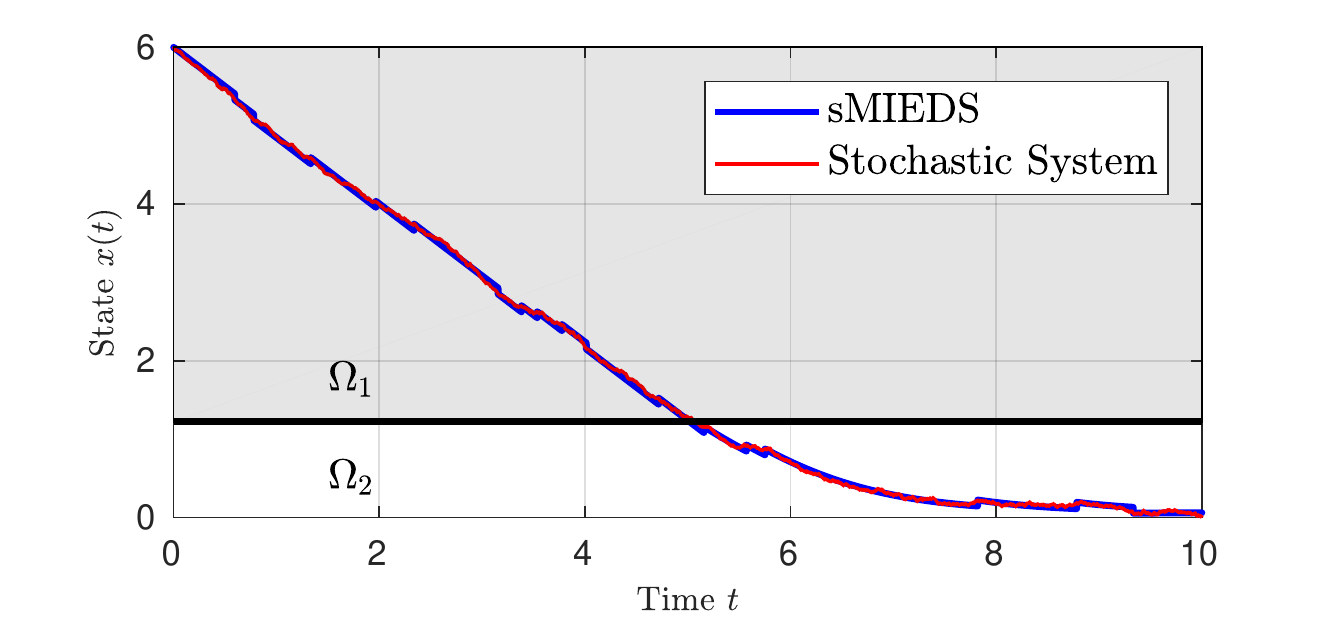}\\
\includegraphics[width=0.4\textwidth]{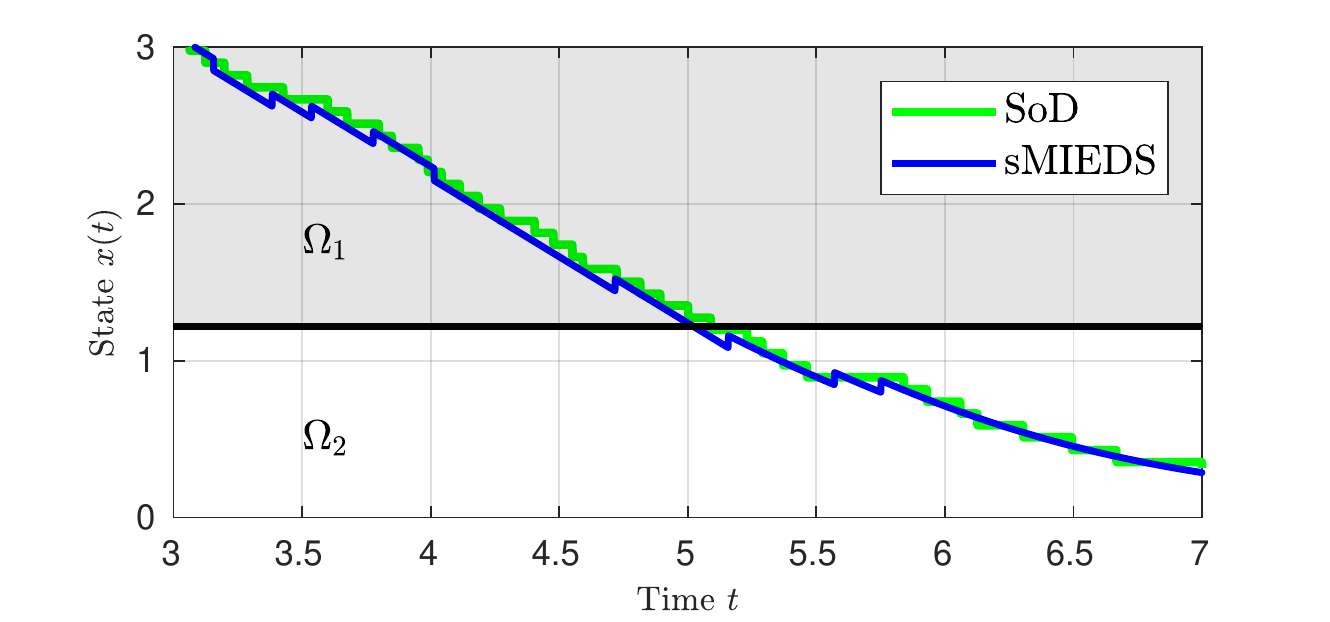}\\
\includegraphics[width=0.4\textwidth]{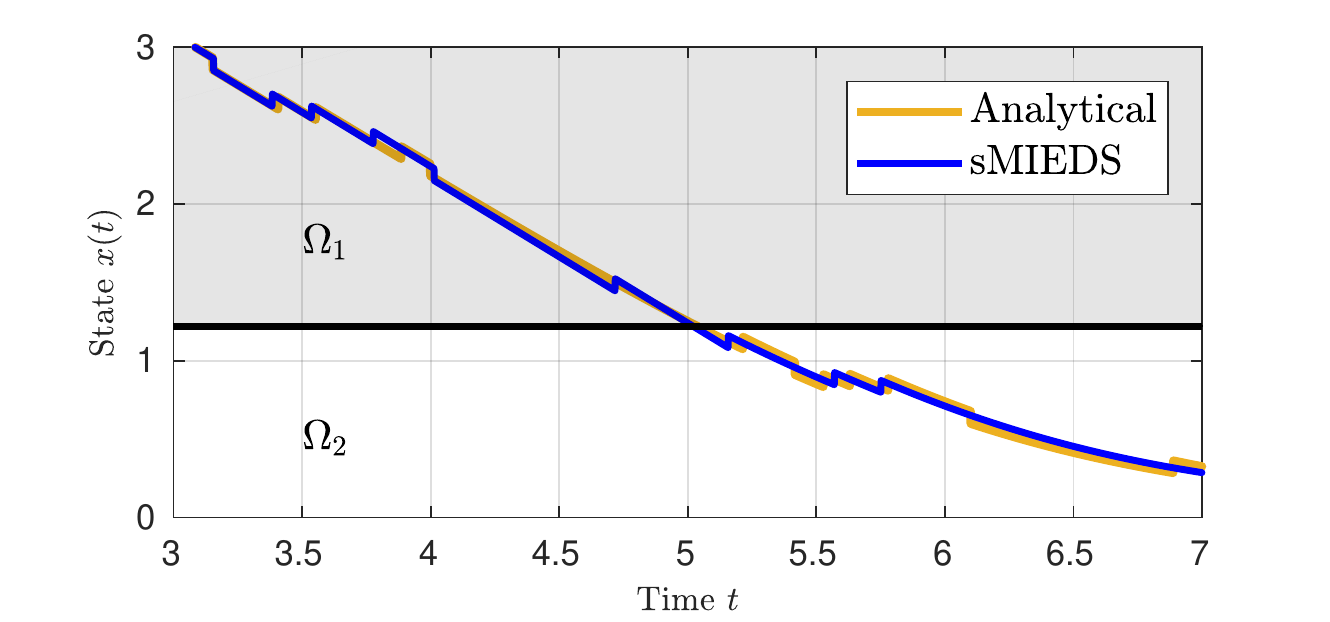}
    \caption{Stochastic example, for which we demonstrate the benefits of our approach. The first plot shows the stochastic trajectory of the system in red and our method in blue.
    Every jump corresponds to communication due to noise effects. The local model is updated when the horizontal line is crossed. In the second plot we see an improvement in communication savings compared to the send-on-delta protocol, which is highlighted by zooming in.
    Furthermore, we show in the third plot similar performance with respect to the analytical model, with the addition benefit of significantly reducing the model complexity.\vspace{-0.2cm}}
\label{fig:tanhStochastic}
\end{figure}
We consider the stochastic system 
$\dot{x}(t) = -\tanh(x(t)) + \frac{1}{10}\epsilon(t)$,
where $\epsilon(t)$ is standard white noise. We assume $x(0) = 6$ and use the Euler-Maruyama method \cite{schuss2009theory} to sample from the system. We consider a fixed time horizon of $T = 10\,\mathrm{s}$ 
and check according to the \mieds algorithm different numbers of partitions of the state space. For the choice of $\lambda = 0.01$, we obtain $m^* = 2$ and hence, two local domains. We determine the switching point by evaluating the solution to $\dot{\bar{x}}(t) = -\tanh(\bar{x}(t))$ in $t=5\,\mathrm{s}$. The local models are obtained with the aid of Taylor expansions and the \mieds algorithm yields a linear function $(k_1=1)$ within $\Omega_1$ and a cubic polynomial $(k_2=3)$ within $\Omega_2$.

In this experiment, we compare performance based on the number of state transmissions due to the trigger $\gamma_{\rm noise}$ in \eqref{eq:noisetrigger}.
In particular, we compare different choices of $\hat{f}$ in Equation \eqref{eq:hatx}: 
\begin{itemize}
\item Analytical predictions $\hat{x}^{\rm Analytical}(t)$, based on the analytical model $f(x) = -\tanh (x)$;
\item \emieds predictions $\hat{x}^{\rm sMIEDS }(t)$, based on the local approximations $\hat{f}^i$;
\item Send-on-delta predictions $\hat{x}^{\rm SoD}$, which use a zero-order hold \cite{Mi06} and therefore, $\hat{f} \equiv 0$.
\end{itemize}
Hence, for every fixed trajectory of the stochastic system $x(t)$, we compute the three predictions $\hat{x}^{\rm Analytical}(t)$, $\hat{x}^{\rm sMIEDS }(t)$, and $\hat{x}^{\rm SoD}(t)$ and enforce the trigger \eqref{eq:noisetrigger} with $\delta_{\rm noise} = 0.075$.
In \fig \ref{fig:tanhStochastic}, we illustrate one trajectory of the system (in red). In the first subfigure, we see  the predictions $\hat{x}^{\rm sMIEDS }(t)$ (in blue). Every jump corresponds to transmitting the current state. 
The other two plots show a zoom in, to yield a direct comparison to the two benchmarks (send-on-delta in green and analytical model in orange). 
Due to the stochastic nature of the system, we run $100$ simulations with the same parameters, initial conditions, and time horizon. Then, we average the number of communication instances and obtain
$$ \gamma_{\rm noise}^{\rm SoD} \approx 78, \quad \gamma_{\rm noise}^{\rm Analytical} \approx 15 ,\quad \gamma_{\rm noise}^{\rm \emieds } \approx 16. $$
Clearly, the SoD-architecture requires significantly more communication and demonstrates that for this example, our algorithm saves communication by a factor of 5. On the other hand, the performance of the analytical model is very comparable to our algorithm. Hence, we show that we are close to optimal in terms of the number of communication instances, while decreasing model complexity significantly. The cost of transmitting a model is not considered here and would result in $4 = k_1 + k_2$ additional weights that need to be send. In the beginning, for the initial model, and when the black line separating $\Omega_1$ from $\Omega_2$ is crossed. For the SoD protocol this would not be necessary.

\section{Discussion}
Using ideas from the MDL principle, the \mieds algorithm is developed herein and proposed as a novel concept to optimally encode dynamical systems. 
We empirically show that the optimal encoding happens somewhere between two extremes of a spectrum: either storing the complete state history, or an initial state and the whole dynamics function. The MIEDS algorithm is applied to multi-dimensional deterministic systems to illustrate the core ideas and extended to cope with stochastic systems (\emieds). 
While the method introduces some computational overhead to compute the local approximations, we demonstrate improvements in terms of more efficient representations. In particular, for event-triggered state estimation, we show the net benefit in terms of reduced communication.
This paper is a proof of concept that information from dynamical systems can be efficiently encoded through local dynamics approximations. However, there are there are still many challenges that need to be addressed in future work.

Generalizing the ideas developed herein to high-dimensional stochastic and multi-agent problems is an possible next step. While the main ingredients for this are developed in this paper, demonstrating the benefits (\eg communication savings) for applications at the scale of real-world problems is still to be shown.
Theoretically quantifying the introduced error to state trajectories by approximating the dynamical function is also a topic of ongoing research. 
While we assume the dynamics function $f$ to be known, it could likewise be learned from data.  In particular, for nonparametric techniques, where the learned function is typically represented by a large data set, the proposed techniques are relevant.  

\section*{Supplementary Material}
Matlab implementations are available at: \url{https://github.com/amehrjou/CDC2018}
\bibliographystyle{IEEEtran}      
\bibliography{database}

\end{document}